\documentclass[conference]{IEEEtran}
\usepackage[utf8]{inputenc}
\IEEEoverridecommandlockouts
\usepackage{cite}
\usepackage{amsmath,amssymb,amsfonts}
\usepackage{bm}
\usepackage{setspace}
\usepackage{graphicx}
\usepackage{float}
\usepackage{textcomp}
\usepackage{xcolor}
\usepackage{subcaption}
\usepackage{tabularx}
\usepackage{multicol}
\usepackage{multirow}
\usepackage{footnote}
\usepackage{tabularx}
\usepackage[flushleft]{threeparttable}
\usepackage[ruled,vlined,noresetcount]{algorithm2e}

\usepackage{algorithm}
\usepackage{algpseudocode}



\def\BibTeX{{\rm B\kern-.05em{\sc i\kern-.025em b}\kern-.08em
    T\kern-.1667em\lower.7ex\hbox{E}\kern-.125emX}}
\begin{document}

\title{Comparative Analysis of Learning-Based Methods for Transient Stability Assessment\\
\thanks{This work was supported partially by the Natural Sciences and Engineering Research Council (NSERC) Grant, NSERC 2019-05336, and ARL W911NF1910110, and partially by the Fonds de Recherche du Quebec-Nature et technologies under Grant FRQ-NT PR-298827.}
}

\small
\author{
    \IEEEauthorblockN{
     Xingjian Wu\textsuperscript{1}, 
     Xiaoting Wang\textsuperscript{2},
        Xiaozhe Wang\textsuperscript{1}, 
        Peter E. Caines\textsuperscript{1}, 
        Jingyu Liu\textsuperscript{1}       }
    \IEEEauthorblockA{
       \textsuperscript{1}\textit{Department of Electrical and Computer Engineering}, \textit{McGill University}, Montreal, Canada\\
        \textsuperscript{2}\textit{Department of Electrical and Computer Engineering}, \textit{University of Alberta}, Edmonton, Canada\\
       xingjian.wu@mail.mcgill.ca;
        xiaotin5@ualberta.ca;
       xiaozhe.wang2@mcgill.ca;
       peterc@cim.mcgill.ca
    }
    }
\normalsize

\maketitle

\begin{abstract}
Transient stability and critical clearing time (CCT) are important concepts in power system protection and control. 
This paper explores and compares various learning-based methods for predicting CCT under uncertainties arising from renewable generation, loads, and contingencies. Specially, we introduce new definitions of transient stability (B-stablilty) and CCT from an engineering perspective. For training the models, only the initial values of system variables and contingency cases are used as features, enabling the provision of protection information based on these initial values. To enhance efficiency, a hybrid feature selection strategy combining the maximal information coefficient (MIC) and Spearman’s Correlation Coefficient (SCC) is employed to reduce the feature dimension. The performance of different learning-based models is evaluated on a WSCC 9-bus system.

\end{abstract}


\section{Introduction}
\color{black}Transient stability refers to the capability of the system's synchronous machines to remain synchronized after large disturbances\cite{gomez2018electric}, which is critical in power system protection and control. \color{black} Critical clearing time (CCT) is a critical concept in transient stability, which denotes the maximum fault clearing time (FCT) at which the system could remain in synchronism. 
\color{black} Traditionally, time domain simulation (TDS) is widely used for CCT determination via a bisection algorithm \cite{7225947}. While Monte Carlo simulations based on TDS address power system uncertainties in transient stability analysis, their heavy computational burden limits application \cite{faried2010probabilistic}. Surrogate modeling (polynomial chaos \cite{Liu2022}, neural network (NN) \cite{6713818,shi2024online})-based methods are applied for fast and accurate CCT prediction. Specially, various NN-based models (e.g., multi-layer perception (MLP), general regression NN (GRNN), adaptive neuro-fuzzy inference systems (ANFIS)) have been adopted for CCT estimation in \cite{6713818}, considering different load conditions and contingency. 

Recently, Shi et al. employed a convolutional neural network (CNN) to predict CCT, demonstrating the effectiveness of one-dimensional CNNs for CCT regression \cite{shi2024online}. However, CNN models require extensive parameter training, requiring a large dataset, and substantial computational power. MLP presented strong regression performances in CCT estimation while only considering a particular contingency \cite{KARAMI2013279}. Furthermore, most of these NN-based models used lack physical/mathematical interpretability.  In this regard, Liu et al. proposed a groundbreaking Kolmogorov-Arnold Networks (KAN) model with fewer layers and nodes, achieving comparable results to MLP models across many tasks \cite{liu2024kan}. This approach also improved interpretability by incorporating learnable activation functions in network edges. Therefore, this paper explores the use of the KAN model in CCT prediction. To the best of our knowledge, it is the first time to apply the KAN method for CCT estimation. 

The high dimensionality of features is another key issue hindering the application of NN-based models in CCT prediction. 
To tackle this issue, Mei et al. first proposed using random forests to extract key features, and the Pearson correlation matrix to eliminate redundant features \cite{mei2022modeling}. A similar combined method of XGBoost and correlation matrix was applied in \cite{8620201}.  However, the Pearson coefficients are limited to linear correlations. In contrast, the Spearman Correlation Coefficient (SCC) is proposed for its strength in handling monotonic relationships, whether linear or nonlinear, between paired data \cite{9173790}.  Maximum Information Coefficient (MIC) is another alternative for feature selection with low computational complexity and high efficiency \cite{reshef2011detecting}. Given the salient properties of MIC and SCC, this paper employs a hybrid feature selection (MIC+SCC) strategy to pinpoint key features for training CCT prediction models. 

However, even with an advanced feature selection strategy, it is essential to determine the original feature pools. All state and algebraic variables at each time instant can be used as features, while 
they may not be instantly accessible in real-time.
Chen et al. selected static features, which enhanced the transient stability analysis by reducing the burden on electrical devices for tracking variables and by providing essential protection information \cite{8881153}. Therefore, considering hardware constraints and protection goals and to address the aforementioned challenges, this paper will apply and compare different learning-based models including NN-based methods like MLP, ANFIS, GRNN, CNN, KAN, and shallow machine models (e.g., linear regression, decision tree, random forest, and XGBoost) for CCT predictions. Specially, the main contributions are as follows.
\begin{itemize}
    \item A specific definition for transient stability, called B-stability, is adopted, further providing a clearer definition of CCT.
     \item Only the initial values of rotor angle, load power, generator power, and contingency are selected as features for CCT estimation, enabling immediate protection instructions once initial states are determined.
    \item A hybrid feature selection algorithm (MIC+SCC) is further integrated to reduce computation complexity and enhance efficiency, revealing robust and standard traits to select key features.
    \item Comparative studies of using learning-based methods, especially, various NN models for CCT prediction are conducted to verify their efficiency, accuracy, and scalability from an engineering perspective.  
\end{itemize}

 The rest of this paper is organised as follows. Section  II introduces the mathematical formulation of the CCT problem. Section III presents the learning-based framework for CCT estimation. Section IV shows the simulation results and discussion. Section V gives the conclusion.




 \color{black}

\color{black}
\section{Problem Formulation for CCT Prediction}
\subsection{Power System Dynamic Modeling Under Uncertainty} \color{black}
\color{black}The power system dynamic model can be represented by a set of differential equations given below \cite{Liu2022}: 
\color{black}
\begin{subequations}
\label{eq:dynamic}
\begin{align}
    \dot{\bm{x}}&=\bm{f}\left (\bm{x},\bm{y},\bm{u},\bm{\xi} \right )  \label{eq:dynamic1_un} \\
     \bm{0}&=\bm{g}\left (\bm{x},\bm{y},\bm{\xi} \right ) \label{eq:dynamic2_un}
\end{align}
\end{subequations}
\color{black}
where $\bm{x}$ are the generation state variables, $\bm{y}$ are algebraic variables (e.g., bus voltages magnitudes and angles), 
and $\bm{u}$ denote the system' discrete inputs (e.g, fault occurrence and switching operation of tap changers). 
$\bm{\xi}$ denote uncertainty sources (e.g., solar power, load power). 
$\bm{f}(\cdot)$ describe the dynamics of the state variables. 
$\bm{g}(\cdot)$ is the set of algebraic equations, 
describing internal relationship between different variables and power balance. 

This paper aims to analyze the transient stability of power systems under large disturbance. It is assumed that the system operates normally from the initial time $t_{0}$ until a fault occurs at $t_{1}$. After some time, the system clears the fault at $t_{2}$. Thus, the topology of the network changes at $t_{1}$ and $t_{2}$, respectively,  with new dynamic systems instantaneously. To assess transient stability, we propose B-stable index in the following section.  
\color{black}


\subsection{Boundary Stable (B-stable) and CCT}
TDS depicts trajectories of state variables in the time domain. 
The transient stability can be determined by comparing the maximum rotor angle difference between any two generators 
during the simulation \color{black}\cite{ZHOU2019379}\color{black}. If the maximum difference is less than 360 degrees, then the system is regarded as a transient stable case and all generators are seen to remain synchronism after the large disturbance. Otherwise, the system is seen as a transient unstable system. 
The B-stable is also defined in terms of the generators' rotor angle deviations.

\noindent \textbf{Definition 1}: The \textit{\textbf{B-stable index}} $\eta_{t}$ for a multi-machine system with a set of generators $S$ at time instant $t$ is defined as: 
\begin{equation}
\setlength{\abovecaptionskip}{2pt}
\setlength{\belowcaptionskip}{2pt}
\eta_{t}=\left \{ \left | \delta_{i,t}-\delta _{j,t} \right | \; | \; i,j\subseteq S,i\ne j,t\ge t_{1}  \right \}_{max}. 
\end{equation}

\noindent \textbf{Definition 2}: For a multi-machine system, it is said to be \textit{\textbf{B-stable}} at the level $\beta$ if $\eta_{t}$ satisfies the equation (\textbf{\ref{eq:bstable beta}}) at any time $t\ge t_{1}$: 
\begin{equation}
\setlength{\abovecaptionskip}{2pt}
\setlength{\belowcaptionskip}{2pt}
    \eta_{t}-\beta \leqslant 0.
    \label{eq:bstable beta}
\end{equation}
For a \textit{\textbf{B-stable}} system, generators are regarded to remain synchronism, which further determines the transient stability of the system at the level $\beta$.

\noindent \textbf{Definition 3}: The \textbf{\textit{critical clearing time}} (CCT)  
$T_{\mathrm{cct}}$ \color{black} is defined as the maximum fault duration time such that the \textit{\textbf{B-stable }index} $\eta_{t}$ satisfies equation (\textbf{\ref{eq:bstable beta}}) at any time $t\ge t_{1}$. 

\noindent \textbf{Remark 1}. Some works set the boundary value $\beta$ to $360^{\circ}$ \cite{8920121}, 
while this paper selects  $\beta=180^{\circ}$ to make the system more conservative. 

\begin{figure}[]
\setlength{\abovecaptionskip}{0.02cm}
\setlength{\belowcaptionskip}{0.02cm}
  \centering
  \begin{subfigure}[b]{0.495\linewidth}
\includegraphics[width=\linewidth]{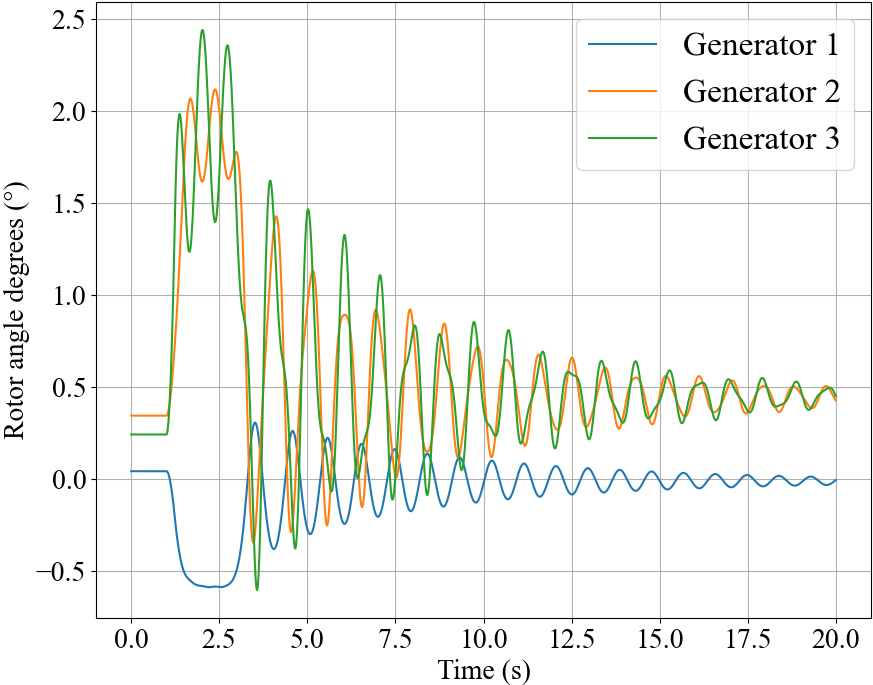}
    \caption{FCT=1.238017 s}
    \label{fig:stable example 1.1}
  \end{subfigure}
  \begin{subfigure}[b]{0.49\linewidth}
    \includegraphics[width=\linewidth]{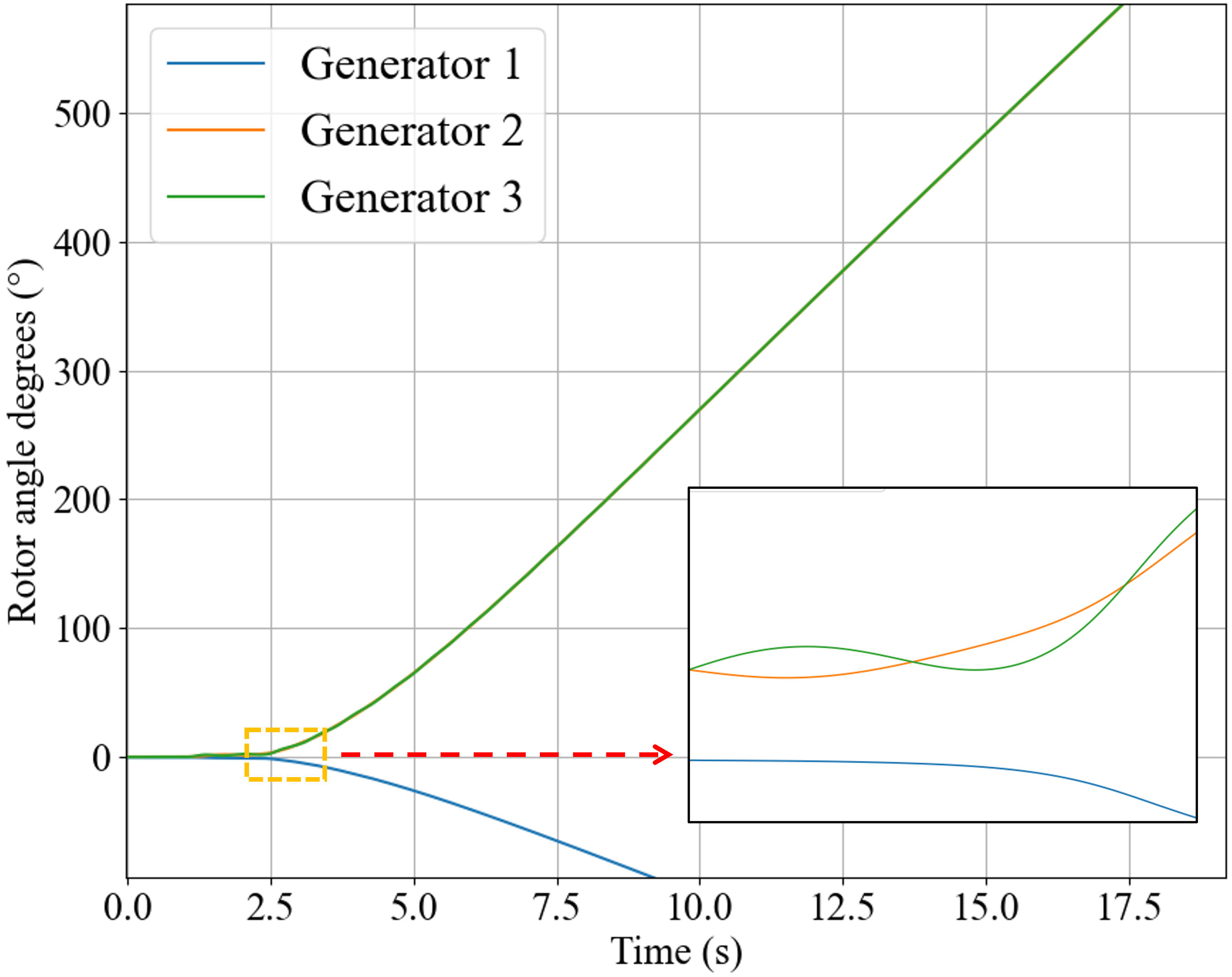}
    \caption{FCT=1.238213 s}
    \label{fig:unstable example 1.1}
  \end{subfigure}
  \caption{Rotor angle curves of determining B-stability ($t_{1}$=1 s)}
  \label{fig:delta curve 9}
 \vspace{-0.2in} 
\end{figure}

Fig. \ref{fig:delta curve 9} shows two scenarios of CCT search by TDS for a WSCC 9-bus system. Rotor angle curves depict trajectories of state variable $\delta$ for each generator. A three-phase ground fault occurs at Bus 9 at $t_{1}=1$ second and is cleared by opening line 6–9. The system remains B-stable if FCT is 1.238017 seconds but fails if FCT is 1.238213 seconds. 

\begin{figure*}[]
\centerline{\includegraphics[width=1\textwidth]{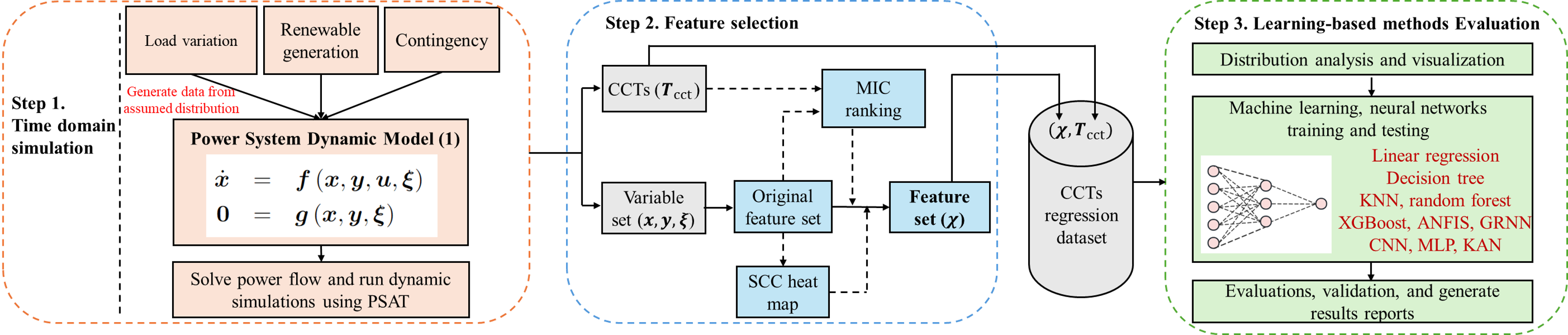}}
\caption{The implementation details of the learning-based framework for CCT estimation}
\label{fig:flow chat 1}
\vspace{-0.1in}
\end{figure*}
%

\section{The Learning-based Framework for CCT Prediction}
%
For transient stability assessment, CCT is typically determined using TDS 
\cite{7225947}\color{black}. However, rapid changes in renewable generation, load variation, and the increasing probability of severe faults due to network complexity necessitate fast CCT evaluation. Therefore, this paper implements and compares different learning-based methods for fast CCT estimation. Furthermore, feature section algorithms are applied to select key features and reduce input dimension (see Section \ref{sec:feature}).
\subsection{Overview of the Framework}
Fig. \ref{fig:flow chat 1} illustrates the implementation details of different learning-based methods for CCT prediction, which includes three consecutive steps.
First, carry out TDS using \eqref{eq:dynamic} to obtain CCTs for all generated scenarios (samples of $\bm{\xi}$) and specific variables will be captured in this step to construct a variable set. Pass the variable set and CCTs to Step 2.
Specially, a hybrid feature selection algorithm, namely,  maximal information coefficient (MCC)+Spearman correlation coefficient (SCC), will be applied to dimension reduction by selecting key features from the original feature pool. Finally, train and compare different basic machine learning models and state-of-the-art neural network tools for CCTs prediction. 

\noindent \textbf{Remark 2}. In Step 1, we generate input data for load and renewable generation power based on the assumed distribution using the UQLab toolbox \cite{lataniotis2015uqlab}. The corresponding outputs include CCTs and a variable set. This variable set contains all state variables $\bm{x}, \bm{y}$ and uncertainty sources $\bm{\xi}$ after solving the power flow at the initial time $t_0$. Specially, the bisection algorithm \cite{7225947} is used to search for the CCTs based on TDS through PSAT/MATLAB \cite{milano2004power}.
\vspace{-0.1in}

\color{black}

\subsection{Feature selection} \label{sec:feature} 
While all state and algebraic variables at each time instant can be used to predict CCT, 
they may not be accessible instantaneously in real-time. 
Moreover, a large number of features can lead to computational complexity and high memory usage.   Therefore, 
only the initial values of rotor angle $\delta$, generator active power $P_{G}$, reactive power $Q_{G}$, load power $P_{d}$, and contingency cases are selected for the original feature pools. These features will be further processed by the MIC+SCC strategy. Table \ref{tab:13 original features} presents an example of original feature pools for the WSCC 9-bus test system. \color{black} 
 

\begin{algorithm}
\small 
\begin{algorithmic}[1]
\State Calculate \textbf{MIC} values between each feature with the target \textbf{(CCT)} by \eqref{eq:MIC1}-\eqref{eq:MIC2}.
\State Rank and list MIC values based on step 1.  
\State Delete variables that have low MIC $\le$ 0.1.  
\State Calculate \textbf{SCC} values between each pair of features using \eqref{eq:SCC}, and depict the \textbf{heat map}. 
\State Find the cell SCC $ \ge$ 0.5 \textbf{or} SCC $\le$ -0.5, delete the variable with lower MIC ranking (calculated using \eqref{eq:MIC1}-\eqref{eq:MIC2} in step 1). 
\State Concatenate remaining variables with CCT and generate \textbf{($\chi,T_{\mathrm{cct}}$)} set for training.
\end{algorithmic}
\caption{MIC+SCC strategy for feature selection}
\label{al:algorithm}
\end{algorithm}
\normalsize
\noindent \textbf{MIC+SCC Feature Selection Algorithm:} 
This paper uses a hybrid feature selection method combining MIC and SCC, summarized in \textbf{Algorithm \ref{al:algorithm}}.  
MIC, proposed by Reshef \textit{et al.}, is widely used for its low computational complexity and high accuracy \cite{chen2023eeg}. \color{black} The MIC of two vectors (e.g., $i^{th}$ feature $\chi_{o}^{i}$ 
 of original feature set $\chi_{o}$ and $T_{\mathrm{cct}}$) can be calculated using \cite{zhang2014novel}:
\small
\begin{eqnarray}  
\label{eq:MIC1}
&& \mathrm{MIC}(\chi_{o}^{i},T_{\mathrm{cct}})=\max\frac{I(\chi_{o}^{i},T_{\mathrm{cct}})}{\log_{2}{\min \left \{ n_{\chi_{o}^{i}},n_{T_{\mathrm{cct}}} \right \} } }\\
&& I(\chi_{o}^{i},T_{\mathrm{cct}}) = H(\chi_{o}^{i})+H(T_{\mathrm{cct}})-H(\chi_{o}^{i},T_{\mathrm{cct}}) \label{eq:MIC2} \\
&&= \sum_{j=1}^{n_{\chi_{o}^{i}} }p(\chi_{o}^{i,j})\log_{2}{\frac{1}{p(\chi_{o}^{i,j})}}+ \sum_{k=1}^{n_{T_{\mathrm{cct}}}}p(T_{\mathrm{cct}}^{k})\log_{2}{\frac{1}{p(T_{\mathrm{cct}}^{k})}} \notag\\
&& -\sum_{j=1}^{n_{\chi_{o}^{i}}} \sum_{k=1}^{n_{T_{\mathrm{cct}}}}p(\chi_{o}^{i,j},T_{\mathrm{cct}}^{k})\log_{2}{\frac{1}{p(\chi_{o}^{i,j},T_{\mathrm{cct}}^{k})}} \notag
\end{eqnarray}
\normalsize
with $ n_{\chi_{o}^{i}} \times n_{T_{\mathrm{cct}}} < N^{0.6}$.
$n_{\chi_{o}^{i}}$ and $n_{T_{\mathrm{cct}}}$ are numbers of bins that separate $\chi_{o}^{i}$ and $T_{\mathrm{cct}}$. $N$ is the number of samples. $p(\chi_{o}^{i,j})$ is the probability of $\chi_{o}^{i}$'s value that fall in $j^{th}$ bin, $p(T_{\mathrm{cct}}^{k})$ is the probability of $T_{\mathrm{cct}}$'s samples that fall in $k^{th}$ bin, and  $p(\chi_{o}^{i,j},T_{\mathrm{cct}}^{k})$ is the probability that $\chi_{o}^{i}$ falls in $j^{th}$ bin and $T_{\mathrm{cct}}$ falls in $k^{th}$ bin simultaneously.   
The ranking of MIC values reflects the information each feature provides to the target ($T_{\mathrm{cct}}$), facilitating feature selection, \color{black} i.e., the features with the highest MIC values are selected. 

SCC quantifies the strength of a monotonic relationship between two paired data (e.g., $\delta_{1}$ and $P_{G_{1}}$), which can be calculated through: 
\begin{align}
    \mathrm{SCC}_{\delta_{1},P_{G_{1}}}=1-\frac{6\sum_{i=1}^{n}(R(\delta_{1,i})-R(P_{G_{1},i}))^{2} }{n\left ( n^{2}-1 \right ) } \label{eq:SCC}
\end{align}
where $n$ is the number of observation pairs and $R(\cdot)$ is the rank of the sample in the series. SCC is 
\color{black}  generally regarded as a robust detector for nonlinear correlations \cite{9173790}. 
The SCC ranges from -1 to 1, indicating a stronger correlation as the values increase. Typically, an absolute SCC greater than $0.5$ indicates a strong or moderate correlation between two variables \cite{mukaka2012guide}. Consequently, this paper uses 0.5 as a threshold for correlation. \color{black} The hybrid selection algorithm leverages both methods to eliminate features based on their contributions to the targets ($T_{\mathrm{cct}}$), preventing the elimination of all highly correlated feature pairs. \color{black} After applying this method, the refined feature set is sent to training. 

\vspace{-0.1in}

\subsection{Evaluation}
To evaluate each model, we use $r^{2}$, MSE (Mean squared error), MAE (Mean absolute error), and MAPE (Mean absolute percentage error). These metrics are given by \cite{10301647} \cite{kim2016new}: 
\small
\begin{eqnarray}
  &r^{2}=1-\frac{\sum_{i=1}^{N}\left ( T_{\mathrm{cct}}^{(i)}-\widehat{T}{_{\mathrm{cct}}^{(i)}}  \right )^{2}   }{\sum_{i=1}^{N}\left ( T_{\mathrm{cct}}^{(i)}-\bar{T}_{\mathrm{cct}}   \right )^{2}},  \mathrm{MSE}=\frac{1}{N} \sum_{i=1}^{N}\left ( T_{\mathrm{cct}}^{(i)}-\widehat{T}{_{\mathrm{cct}}^{(i)}}   \right )^{2}
    \label{eq:metric_eval1} \\
    &\mathrm{MAE}=\frac{1}{N} \sum_{i=1}^{N}\left |T_{\mathrm{cct}}^{(i)}-\widehat{T}{_{\mathrm{cct}}^{(i)}}   \right |,  \mathrm{MAPE}=\frac{1}{N} \sum_{i=1}^{N}\left | \frac{T_{\mathrm{cct}}^{(i)}-\widehat{T}{_{\mathrm{cct}}^{(i)}}  }{T_{\mathrm{cct}}^{(i)}}  \right |    \label{eq:metric_eval2}
\end{eqnarray}
\normalsize
where $N$ is the total amount of samples, $T_{\mathrm{cct}}^{(i)}$ is the $i^{th} $ true value, $\widehat{T}{_{\mathrm{cct}}^{(i)}} $ is corresponding predicted value and $\bar{T}_{\mathrm{cct}}$ is the sample mean value. 
\captionsetup{
  font=footnotesize,
  justification=centering,
  singlelinecheck=false
}

\begin{table}[htbp]
\setlength{\abovecaptionskip}{0.cm}
\centering
\renewcommand{\arraystretch}{1.1}
\caption{ORIGINAL FEATURE POOL FOR THE WSCC 9-BUS TEST SYSTEM}
\resizebox{.99\columnwidth}{!}{
\begin{tabular}{cclc}
\hline
Feature & Time & \multicolumn{1}{c}{Descriptions}     & Amount   \\ \hline
$No.$ & - & Contingency case no. for $i^{th}$ fault,\  $i$=1,2, $\cdots$,10       & 1       \\
$\delta_{i}$ (\textdegree)  & $t_{0}$        & Rotor angle of $i^{th}$ generator,\ $i$=1,2,3  & 3 \\
 $P_{Gi}$ (p.u.) & $t_{0}$      & Active power of $i^{th}$ generator,\ $i$=1,2,3  & 3 \\
 $Q_{Gi}$ (p.u.)  & $t_{0}$        & Reactive power of $i^{th}$ generator,\ $i$=1,2,3  & 3\\
 $P_{di}$ (p.u.)  & $t_{0}$      & Active power of $i^{th}$ load, \ $i$=1,2,3      & 3  \\ \hline
\end{tabular}}
\label{tab:13 original features}
\end{table}
\section{Results and Discussion}
\subsection{Test system and data preparation}
In this section, we evaluate and compare different learning methods on a WSCC 3-machine and 9-bus test system. 
The system includes three loads, each modeled as a Gaussian distribution ranging from 80\% - 120\% of their base power \cite{Liu2022} \cite{4484955}. Additionally, one solar generator and one wind generator are added at Bus \{2,3\}, respectively. The solar generator's active power follows a Beta distribution between 75\% and 115\% of the benchmark, while the wind generator's active power variation follows a Weibull distribution with a 15\% variance \cite{9242254}. Both renewable generators are assumed to have a unity power factor. Parameters for generating random input data (load and renewable generator power) are taken from \cite{Liu2022}, \cite{9242254}.
\begin{table}[htbp]
\small
\setlength{\abovecaptionskip}{0.cm}
\centering
\renewcommand{\arraystretch}{1.0}
\centering
\caption{FAULT CONTINGENCY CONFIGURATION AND STATISTICS OF CCTS BASED ON SAMPLES \color{black} FOR EACH CONTINGENCY}
\label{tab:table label-cct}
\resizebox{.99\columnwidth}{!}{
\begin{tabular}{ccccc} 
\hline
Cont. no. & Location & Remove line        & Mean & STD     \\ 
\hline
1               & Bus 5              & 5$\sim$7                   & 0.301    & 0.0943  \\
2               & Bus 4              & 4$\sim$5                   & 0.314    & 0.0661  \\
3               & Bus 9              & 6$\sim$9                  & 0.214    & 0.0471  \\
4               & Bus 8              & 7$\sim$8                   & 0.264    & 0.0472  \\
5               & Bus 9              & 8$\sim$9             & 0.239    & 0.0366  \\
6               & Bus 6              & 6$\sim$9   & 0.360    & 0.0892  \\
7               & Bus 5              & 4$\sim$5   & 0.356    & 0.0592  \\
8               & Bus 8              & 8$\sim$9      & 0.290    & 0.0424  \\
9               & Bus 6              & 4$\sim$6       & 0.388    & 0.0716  \\
10              & Bus 4, 6           & 4$\sim$6, 6$\sim$9    & 0.284    & 0.0772  \\
\hline
\end{tabular}}
\end{table}
\normalsize

Specially, we consider 10 contingent cases, as shown in Table \ref{tab:table label-cct}. For each case, the fault occurs at $t_{1}=1$s. For each contingency, 150 samples of load and renewable generator power are generated. The sample mean and standard deviation for each contingency case are presented in Table \ref{tab:table label-cct}. For each contingency case, 150 samples are generated. Excluding zero and infinite CCTs, 1489 samples are valid to construct the dataset.

\color{black}

\subsection{Feature Selection and Performance Evaluation}
\color{black}

\color{black}
\begin{figure}[]
\setlength{\abovecaptionskip}{2pt}
\setlength{\belowcaptionskip}{2pt}
\centering
\centerline{\includegraphics[width=0.46\textwidth]{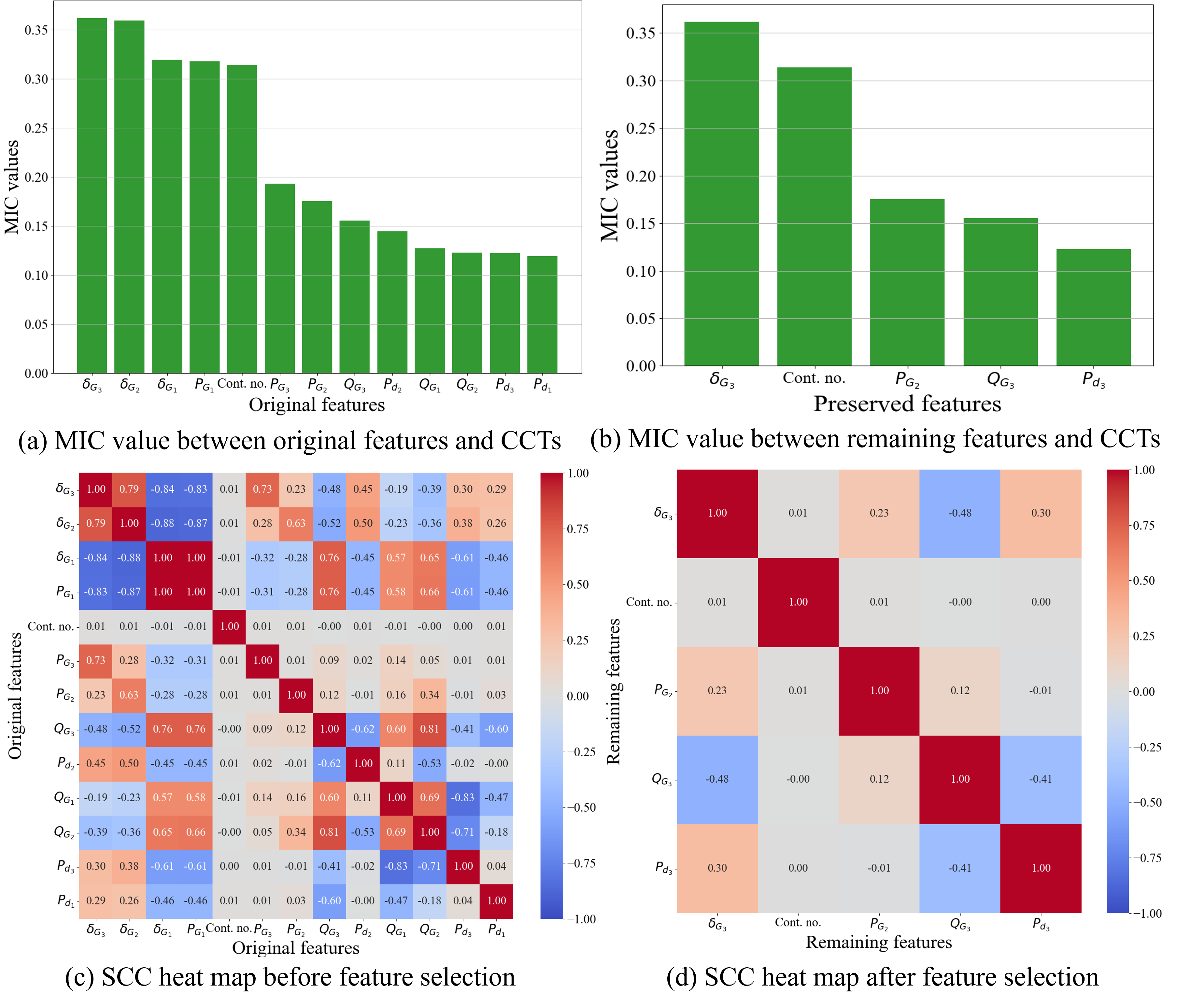}}
\caption{MIC and SCC heat map before and after feature selection. Left: before. Right: after.}
\label{fig:MIC+SCC}
\vspace{-0.2in}
\end{figure}

\noindent \textbf{Feature selection:} As discussed in Section  \ref{sec:feature}  (see Table \ref{tab:13 original features}), the original feature pool contains 13 features. After applying the MIC+SCC algorithm, 5 key features are selected shown in Fig. \ref{fig:MIC+SCC} (b) and (d). The MIC values and SCC heat map before and after feature selection are presented in Fig. \ref{fig:MIC+SCC}. Features are ranked by MIC values, with the heat map highlighting highly correlated features in deep red and blue cells. For example, the SCC value between $P_{G_{1}}$ and $\delta_{G_{1}}$ is 1.00 (Fig. \ref{fig:MIC+SCC} (c)), indicating a strong proportional relationship. Since $P_{G_{1}}$ has a lower MIC value than $\delta_{G_{1}}$, it is removed from the feature pool. However, both $P_{G_{1}}$ and $\delta_{G_{1}}$ are ultimately dropped due to their negative correlation with $\delta_{G_{3}}$, which has the highest MIC value. Once the features are selected, we pass the dataset $(\bm{\chi},\bm{T}_{\mathrm{cct}})$ to train models. 

\color{black}
\noindent \textbf{Comparison of Different Learning-based Models:} We evaluate the performance of different learning-based models for CCT prediction. Table \ref{tab:all models performances} presents the tested models and their evaluation metrics. Specially, to prevent data bias and over-fitting, we employ 5-fold cross-validation technique. Note that most models are evaluated using 5-fold cross-validation, except for CNN and KAN models.  The CNN model uses 8\% of the dataset for validation and 72\% for training, while the KAN model uses 10\% for validation and 70\% for training. More samples are used to train CNN due to its higher number of trainable parameters. Both models share the same test set, comprising 20\% of the dataset.

\color{black}

\begin{table}[!ht]
\setlength{\abovecaptionskip}{0.02cm}
\setlength{\belowcaptionskip}{0.02cm}
\caption{EVALUATION RESULTS OF ML AND NN MODELS}
\label{tab:all models performances}
\centering
\resizebox{.99\columnwidth}{!}{
\begin{tabular}{lccccc} \hline
\multicolumn{1}{c}{Method} & $r^2$ score & MSE & MAE & MAPE[\%] & TET[s]\tnote{*}\\ \hline
Linear regression & 0.511 & 3.43E-03 & 0.050 & 17.57 & 0.06 \\
Decision tree & 0.725 & 1.93E-03 & 0.034 & 11.83 & 0.02 \\
KNN & 0.761 & 1.68E-03 & 0.027 & 9.483 & 0.02 \\
Random Forest & 0.750 & 1.76E-03 & 0.033 & 11.46 & 0.55 \\
XGBoost & 0.782 & 1.54E-03 & 0.030 & 10.92 & 0.10 \\
MLP & 0.961 & 2.72E-04 & 0.011 & 3.731 & 1.25 \\
ANFIS & 0.944 & 3.93E-04 & 0.014 & 5.039 & 29 \\
GRNN & 0.962 & 2.54E-04 & 0.011 & 3.671 & 6.23 \\
CNN & 0.932 & 5.36E-04 & 0.015 & 5.300 & 57 \\
KAN & 0.981 & 1.52E-04 & 0.009 & 3.296 & 43 \\ \hline
\end{tabular}}
\begin{tablenotes}
\item * \footnotesize{TET denotes the total execution time.}
\end{tablenotes}
\end{table}
\normalsize
As can be seen from Table \ref{tab:all models performances}, linear regression shows the worst performance. Decision tree, KNN, random forest, and XGBoost models show better results with higher $r^{2}$ scores and lower MSE, but they still fall short compared to neural network methods. 

\begin{figure}[]
\centerline{\includegraphics[width=0.5\textwidth]{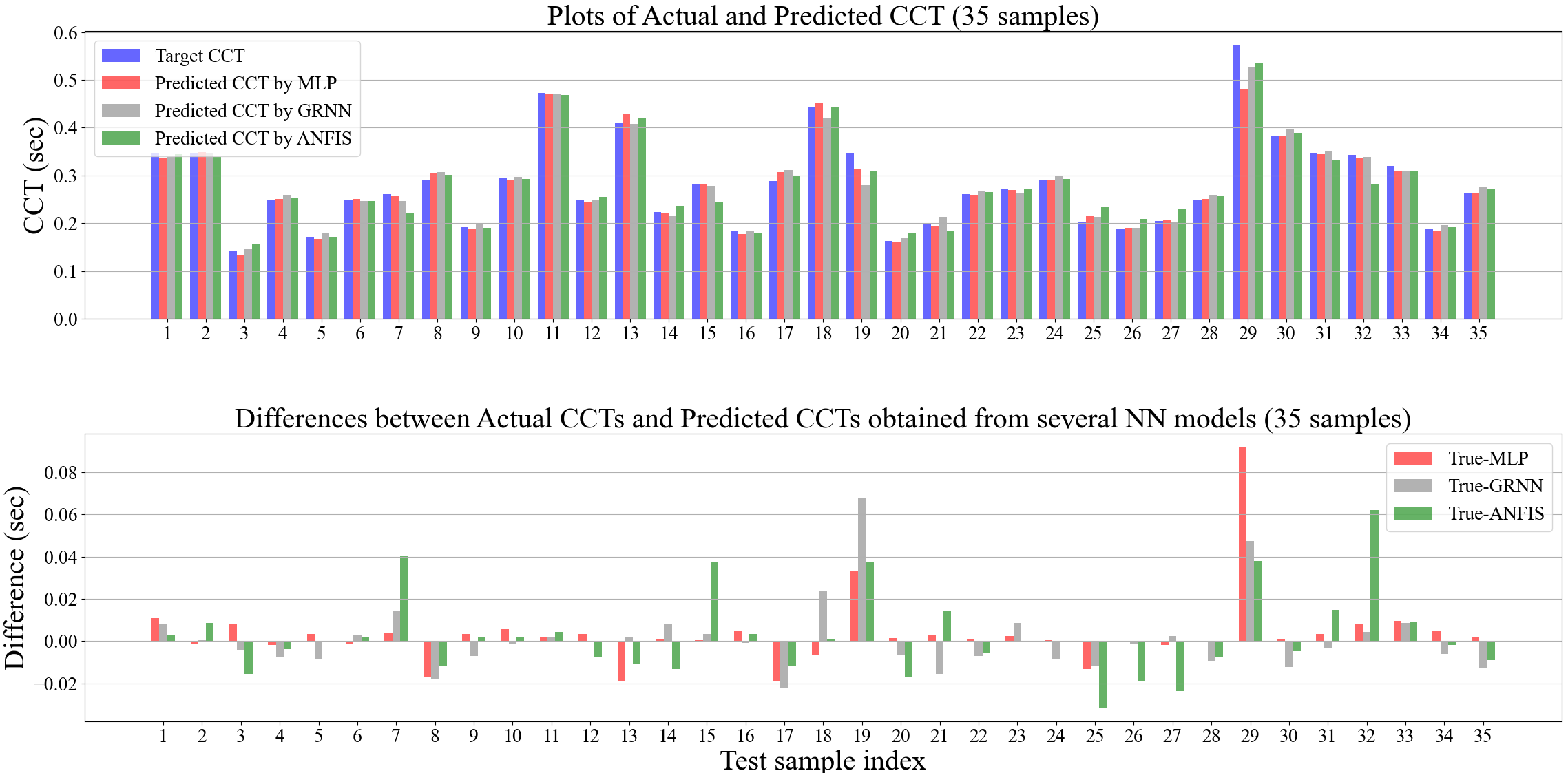}}
\caption{Comparison between real CCTs and predicted CCTs (35 samples)}
\label{fig:3 nn compare}
\end{figure}

A multi-layer fully connected neural network \cite{KARAMI2013279} is built, which has five hidden layers with 15 neurons each, totaling 1066 trainable parameters. Various activation functions (sigmoid, Tanh, ReLu, and logistic) and regularization rates are tested. ReLu with $\alpha$=0.01 shows consistent performance across dataset folds. 
ANFIS \cite{6713818} with a (2-8-2-2-2) structure achieves a lower $r^{2}$ score compared to MLP. GRNN \cite{6713818} shows the second-best regression performance on test sets. However, both ANFIS and GRNN are more time-consuming than MLP. Fig. \ref{fig:3 nn compare} compares the predicted CCT values from MLP, ANFIS, and GRNN models and the ground truth. The lower diagram in Fig. \ref{fig:3 nn compare} illustrates the differences. MLP and GRNN predict accurately for most samples, with differences bounded by $\pm$0.02. However, MLP shows its limitation when predicting extreme values (e.g., a large deviation in $29^{th}$ sample).

\color{black}
CNN \cite{shi2024online} with an architecture comprising 32 filters, a dropout layer, a max-pooling layer, followed by 16 filters, and a fully connected layer with 8 neurons also offers a competitive solution. Leaky ReLU (0.1 leakage) is used for all activation and the original feature set is directly used for prediction due to its internal filtering scheme. However, the CNN model, requiring 2433 trainable parameters, is trained on a dataset of 1050 samples, which is insufficient for optimal performance. Increasing the training sample size may enhance performance. Table \ref{tab:all models performances} shows KAN \cite{liu2024kan} with a structure (5-3-2-1), grid size$=4$ and maximum degree$=3$ outperforms other neural network models in terms of the $r^{2}$ score and all other metrics, corresponding to its exceptional ability in regression. 

\begin{figure}[]
\centering 
\includegraphics[width=0.45\textwidth]{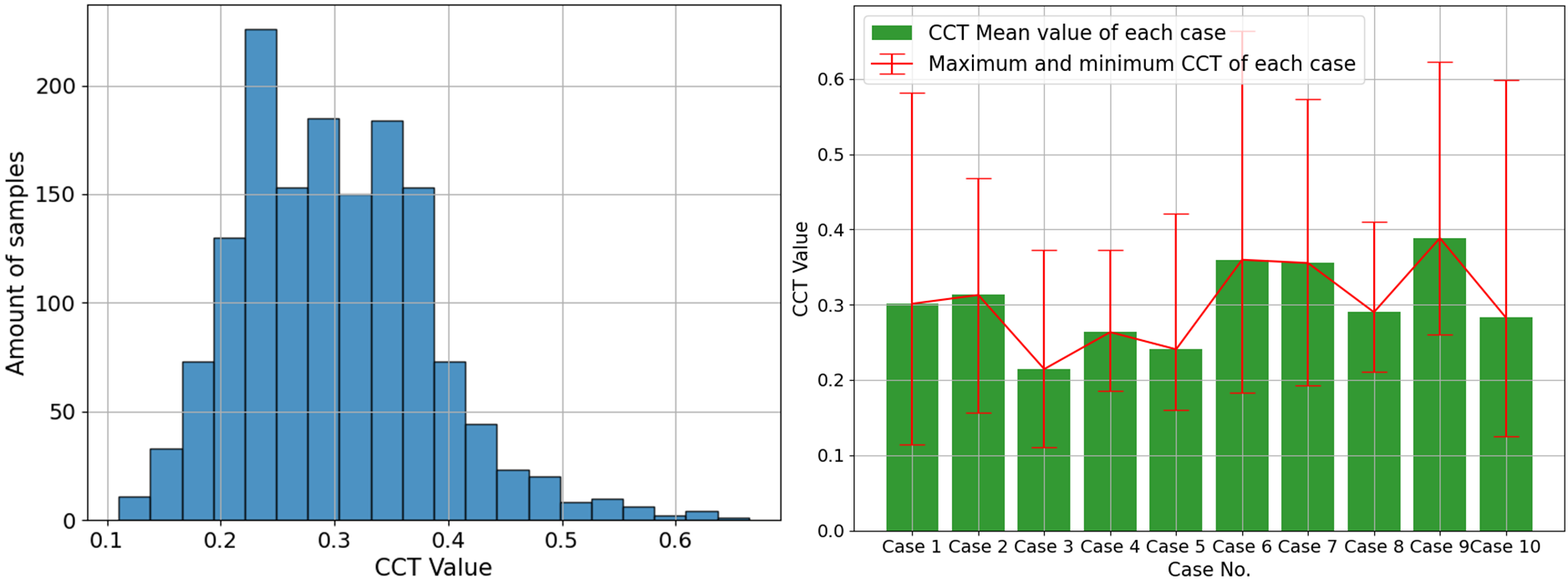}
\caption{The distribution of ground truth CCT (right), and the statistics of CCT vs contingency cases (left) for 1489 samples}
\label{fig:CCT_distribution}
\end{figure}

\noindent \textbf{Evaluation of NN Models for CCT in different ranges:} \color{black}We further evaluate the performance of different NN models across different data clusters as detailed in Table \ref{tab:cct regression distribution}. Fig. \ref{fig:CCT_distribution} depicts the distributions of CCTs (i.e., CCTs obtained from \textbf{Step 1} in Fig. \ref{fig:flow chat 1}) and the relation between CCTs and 10 contingency cases respectively. The target (CCT) is biased towards lower values and exhibits long tails with higher CCTs. Contingency cases significantly influence CCTs (e.g., mean, upper, and lower bounds), indicating that various topological changes impact system stability differently.

We separate three clusters based on CCTs distribution (i.e., $\mathrm{CCTs} \leq 0.25$, $0.25 < \mathrm{CCTs} \leq 0.4$, and $\mathrm{CCTs} >0.4$).  There are fewer samples with CCTs less than 0.25s, which reduces MLP's performance in this range, although it remains high (MSE=9.80E-05). MLP accurately predicts CCTs between 0.25s and 0.4s but loses accuracy for larger CCTs. Accurate predictions for CCTs under 0.25s are crucial, making MLP preferable for CCT prediction. Conversely, ANFIS excels at predicting larger CCTs. GRNN and KAN show consistent performance across both extreme and middle-range CCTs.
\color{black}

\begin{table}[!ht]
\setlength{\abovecaptionskip}{0.02cm}
\setlength{\belowcaptionskip}{0.02cm}
\caption{RESULTS OF NN MODELS W.R.T. CCTS DISTRIBUTION}
\centering
\begin{tabular}{lclclcl}
\hline
NN models & \multicolumn{6}{l}{MSE}                                           \\ \hline
 & \multicolumn{2}{c}{CCT$\le$0.25} & \multicolumn{2}{c}{0.25$<$CCT$\le$0.4} & \multicolumn{2}{c}{CCT$>$0.4} \\ \cline{2-7} 
MLP                & \multicolumn{2}{c}{9.80E-05} & \multicolumn{2}{c}{9.60E-05} & \multicolumn{2}{c}{1.10E-03} \\
GRNN               & \multicolumn{2}{c}{1.76E-04} & \multicolumn{2}{c}{1.62E-04} & \multicolumn{2}{c}{6.16E-04} \\
ANFIS              & \multicolumn{2}{c}{4.06E-04} & \multicolumn{2}{c}{4.00E-04} & \multicolumn{2}{c}{2.76E-04} \\
CNN              & \multicolumn{2}{c}{3.35E-04} & \multicolumn{2}{c}{4.64E-04} & \multicolumn{2}{c}{1.33E-03} \\
KAN              & \multicolumn{2}{c}{1.79E-04} & \multicolumn{2}{c}{1.07E-04} & \multicolumn{2}{c}{2.77E-04} \\
\hline
\end{tabular}
\label{tab:cct regression distribution}
\end{table}
\normalsize

\color{black}

\subsection{Predictions without the knowledge of contingency}

\color{black}
This section evaluates the models without the knowledge of contingency. We observe that all methods listed Table  \ref{tab:all models performances} fail to achieve good prediction results without contingency information, indicating the importance of fault-on topology for CCT regression.  It is challenging to give only initial values from the generators' and loads' sides to make accurate predictions without the knowledge of the specific type of fault. Initial values from generators and loads lack this topological information. 

However, the regression task for CCT prediction is still feasible with additional features at time instant $t_{1}$, even without fault labels. Three generator active power outputs (at $t_{1}$) are added to the original feature pool in Table \ref{tab:13 original features} while remove all fault labels. After applying the MIC+SCC strategy, 6 features remain as shown in Table \ref{tab:added t1}.
Table \ref{tab:performnce after t1} compares the performance of MLP and GRNN, indicating that the CCT prediction can be achieved without pre-knowledge of specific contingency while additional features at time instant $t_1$ are required. These features change instantly at $t_{1}$ due to the sudden topological change, implicitly including topological information. However, the prediction accuracy decreases to an $r^{2}$ of 0.9, demonstrating the limitations of relying solely on these added features.

\color{black}


\normalsize
\begin{table}[!ht]
\setlength{\abovecaptionskip}{0.02cm}
\setlength{\belowcaptionskip}{0.02cm}
\caption{SELECTED FEATURES FOR REGRESSION WITHOUT KNOWING CONTINGENCY}
\centering
\begin{tabular}{cccc}
\hline
Feature          & Time instant & Unit   & MIC   \\ \hline
$\delta_{G_{3}}$ & $t_{0}$      & degree & 0.362 \\
$P_{G_{3}}$      & $t_{1}$      & p.u.   & 0.346 \\
$P_{G_{2}}$      & $t_{1}$      & p.u.   & 0.337 \\
$P_{G_{1}}$      & $t_{1}$      & p.u.   & 0.293 \\
$P_{G_{2}}$      & $t_{0}$      & p.u.   & 0.175 \\
$Q_{G_{3}}$      & $t_{0}$      & p.u.   & 0.155 \\ \hline
\end{tabular}
\label{tab:added t1}
\vspace{-0.12in}
\end{table}
\normalsize



\begin{table}[htbp]
\caption{REGRESSION PERFORMANCES WITH FEATURES IN TABLE. \ref{tab:added t1}}
\centering
\begin{tabular}{ccccc}
\hline
Model & $r^{2}$ score & \multicolumn{1}{c}{MSE} & \multicolumn{1}{c}{MAE} & \multicolumn{1}{c}{MAPE[\%]} \\ \hline
MLP   & 0.892   & 7.49e-4                 & 0.021                  & 8.00                    \\
GRNN  & 0.910   &6.36e-4                 & 0.019                   & 7.08                   \\
\hline
\end{tabular}
\label{tab:performnce after t1}
\end{table}

\color{black}
\vspace{-0.12in}
\section{Conclusion}

This paper applied and compared different learning-based methods for CCT prediction under uncertainty from renewable generation, loads, and contingencies. Specially, we first adopted a B-stability-based approach for defining transient stability and CCT. Only initial values of system variables and contingency cases were selected as features for training different models. This could provide protection information once the initial values are obtained. To further improve efficiency, a hybrid feature selection strategy (MIC+SCC) was integrated to reduce feature dimension. Various learning-based models were evaluated on a WSCC 9-bus system in different scenarios validating their efficiency and scalability. Simulation results demonstrated that: 1) MLP has robust and accurate performance, with fast and convenient traits from an engineering perspective; 2) MLP excels in predicting lower CCT values but is less accurate with larger CCTs; 3) The CNN model's weaker performance could be improved with more training samples; 4) The KAN model with a simple structure, achieves the best regression results on the test dataset. Future studies will involve extending various NN-based methods in large-scale systems. Besides, we will explore input features with extended monitoring time for CCT prediction, without prior knowledge of contingency factors.
%
\color{black}

\bibliographystyle{IEEEtran}
\bibliography{IEEEabrv,xingjian}


\end{document}